\documentclass{article}
\usepackage{epsfig}

\def\n{\noindent}

\begin{document}

\baselineskip .7cm

\author{Navin Khaneja\thanks{Corresponding author: Email:navin@hrl.harvard.edu. 
Division of Engineering and Applied Sciences, Harvard University,
Cambridge, MA 02138. The work was supported by ONR 38A-1077404, AFOSR FA9550-05-1-0443 and AFOSR FA9550-04-1-0427}, 
\ \ Cindie Kehlet\thanks{Centre for Insoluble Protein Structures (inSPIN), Interdisciplinary Nanoscience
Center (iNANO) and Department of Chemistry, University of Aarhus, Denmark}, \ \ 
Steffen J. Glaser\thanks{Department of Chemistry, Technische Universit\"at M\"unchen,
85747 Garching, Germany. This work was supported by the Deutsche Forschungsgemeinschaft, grant Gl 203/4-2.},\ \ 
Niels Chr. Nielsen\thanks{Corresponding author: Email: ncn@chem.au.dk. 
Centre for Insoluble Protein Structures (inSPIN), Interdisciplinary Nanoscience
Center (iNANO) and Department of Chemistry, University of Aarhus, Denmark. The work was supported by the Danish National Research
Foundation, Carlsbergfondet, The Danish Natural Science Research Council, and the Danish Biotechnological Instrumentcentre (DABIC).}}

\vskip 4em

\title{\bf Composite Dipolar Recoupling: Anisotropy Compensated Coherence Transfer in 
Solid-State NMR}

\maketitle

\vskip 3cm

\begin{center} {\bf Abstract} \end{center}

The efficiency of dipole-dipole coupling driven coherence transfer experiments 
in solid-state NMR spectroscopy of powder samples is limited by dispersion of the 
orientation of the internuclear vectors relative to the external magnetic field. 
Here we introduce general design principles and resulting pulse sequences that approach full
polarization transfer efficiency for all crystallite orientations in a powder
in magic-angle-spinning experiments. The methods compensate for the defocusing of 
coherence due to orientation dependent dipolar coupling
interactions and inhomogeneous radio-frequency fields. The 
compensation scheme is very simple to implement as a scaffold (comb) of compensating pulses 
in which the pulse sequence to be improved may be inserted. The degree of compensation 
can be adjusted and should be balanced as a compromise between efficiency and length of the 
overall pulse sequence.
We show by numerical and experimental data that the presented compensation 
protocol significantly improves the efficiency of known dipolar recoupling
solid-state NMR experiment.


\n
\vskip 3em
\newpage
\section{Introduction}

Nuclear magnetic resonance (NMR) spectroscopy is rapidly finding increasingly important 
applications for atomic-resolution structural analysis of biological macromolecules in the 
solid phase \cite{opellanatstruct,griffinnatstruct,oschkinatnat,petkovapnas,griffinpnas1}.
This opens up new avenues for studying "insoluble" protein
structures such as membrane proteins, fibrils, and extracellular matrix proteins which are exceedingly
difficult to analyze using conventional atomic-resolution structure determination methods, including
liquid-state NMR and X-ray crystallography. The 
progress in biological solid-state NMR relies on continuous development of instrumentation,
sample preparation methods, and the underlying NMR methodology. The goal of studying increasingly 
complex molecular systems is a strong motivation for the development of improved solid-state NMR methods. 

The latter challenge motivates the present paper, where we address a fundamental problem of 
coherence transfer in solid-state NMR of "powder" samples. For solids, the internal 
Hamiltonian not only contains isotropic interactions, such as isotropic 
chemical shifts and scalar $J$ couplings, but also anisotropic (i.e., orientation dependent) chemical 
shifts and  dipole-dipole coupling interactions in the case of coupled spin-$1/2$ nuclei. This implies
that each
molecule/crystallite in a "powder" sample may exhibit different nuclear spin interactions 
leading to severe line broadening and thereby reduced spectral resolution and sensitivity. This problem
may be alleviated using magic-angle spinning (MAS), which averages these interactions and hereby results in  
high-resolution conditions for solid samples. However, this also results in loss of useful parts 
of the anisotropic interactions like dipolar couplings, which carry information about distances 
between nuclei and can help in obtaining structural information. 
This has triggered the development of dipolar recoupling techniques 
\cite{dcp,redor,r2,horror,c7,postc7,levittencycl} ,
which selectively reintroduce these couplings to enable measurement of internuclear distances, torsion angles, 
and transfer of magnetization from spin to spin in the molecule. 
Such recoupling experiments are the building blocks of essentially all biological solid-state NMR experiments 
using "powder" samples. However, the orientation dependence 
of the dipolar coupling interaction poses a fundamental challenge for the design of  experiments, e.g. for 
transfer of coherence between spins using dipolar couplings, that are insensitive 
to the orientation dependence. 

The orientation dependency of the dipolar coupling is characterized by Euler angles 
$\beta$ and $\gamma$ expressing the angle between the internuclear axis and the MAS rotor and 
the rotation of the crystallite around the rotor axis, respectively. The discovery of the so 
called $\gamma$-encoded dipolar recoupling by Nielsen and 
coworkers \cite{horror} in 1994, showed that it is possible to eliminate the dependency of coherence transfer 
efficiency on the angle $\gamma$ and increase the transfer efficiency to 73\% from the previously 
accepted maximum of 50\%. In this paper, we map the problem of dipolar recoupling in the presence of 
anisotropies in $\beta$ and the strength of rf field to a problem of control of single spin in 
the presence of rf-inhomogeneity and Larmor dispersion. Using this analogy, we demonstrate how relatively 
simple procedures can combine the concepts of
solid-state NMR $\gamma$-encoded recoupling \cite{griffinnatstruct,redor,r2,horror,c7,postc7,levittencycl}, 
based on coherent averaging methods \cite{haberlenwaugh,scbch,eeht}, with the concepts of
composite pulse sequences \cite{levittfreeman,shaka,levittprog} from liquid-state NMR to make experiments insensitive 
to angle $\beta$. It is now possible to construct a family of dipolar recoupling experiments of 
increasing length and 
degree of compensation that ultimately achieve 100\% transfer efficiency for all orientations of the 
dipolar coupling tensor. 
In practice, the desired level of compensation needs to be 
traded against increased loss of signal due to 
relaxation during the increased pulse sequence duration. 
We show that besides being less sensitive to the orientational dependencies, the compensated 
recoupling experiments are also more robust towards resonance offsets and instrumental artifacts like 
rf inhomogeneity. The general design principle is demonstrated by applications to 
heteronuclear dipolar recoupling and similar applications for homonuclear dipolar recoupling are 
described. Using these compensated pulse sequences, 
we provide experimental performances similar to those obtained using optimal control procedures 
\cite{pontryagin,grape} and recently demonstrated in solid-state NMR \cite{ocdcp,ochorror}. 

\section{Theory}

Consider two coupled heteronuclear spins $I$ and $S$ under magic angle spinning. The spins 
are irradiated with rf fields at their Larmor frequencies along say the $x$ direction. In a 
doubly rotating Zeeman frame, rotating with both the spins at their Larmor frequency, the 
Hamiltonian of the system takes the form 
\begin{equation}
\label{Eq:rotatingH} 
H(t) = \omega_{I}(t) I_z +  \omega_{S}(t) S_z +  \omega_{IS}(t) 2I_zS_z
+ \omega_{rf}^{I}(t) I_x + \omega_{rf}^{S}(t) S_x \quad ,
\end{equation}
where $\omega_{I}(t)$, $\omega_{S}(t)$, and $\omega_{IS}(t)$ represent time-varying chemical shifts for 
the two  spins $I$ and $S$ and the coupling between them, respectively. These interactions may be expressed as
Fourier series $\omega_{\lambda}(t) = \sum_{m=-2}^{2}\omega_{\lambda}^{m}\exp(im\omega_r t)$,
where $\omega_r$ is the spinning frequency (in angular units), while the 
coefficients $\omega_{\lambda}$ $(\lambda = I, S, IS)$ reflect the dependence on the
physical parameters like the isotropic chemical shift, anisotropic chemical shift, the dipole-dipole
coupling constant and through this the internuclear distance \cite{simpson}. 
$\omega_{rf}^{I}(t)$ and $\omega_{rf}^{S}(t)$ are amplitudes of the rf fields on 
spins $I$ and $S$, respectively.  When the rf field strengths 
on the two spins is chosen to be integral (or half integral) multiples 
of spinning frequency, i.e., $\omega_{rf}^{I} = p \omega_r$ and $\omega_{rf}^{S} = q \omega_r$, 
the Hamiltonian for the dipole-dipole coupling in the interaction frame of 
the rf irradiation averages over a rotor period to \cite{gate,idcp} 
\begin{eqnarray}
\label{eq:effmain}
\bar{H}_{IS}= (A_{p+q}^+ Z^{+} -i A_{p+q}^-Y^{+}) + (A_{p-q}^+Z^{-} - i A_{p-q}^-Y^{-}) \quad ,
\end{eqnarray}
where $A_n^{\pm} = \frac{1}{2}(\omega_{IS}^{-(n)} \pm
\omega_{IS}^{(n)})$, $Z^{\pm}= I_zS_z \mp I_yS_y$ and $Y^{\pm} = I_yS_z \pm I_zS_y$. For convenience,
we also define the operator  $X^{\pm} = \frac{1}{2}(I_x \pm S_x)$, which completes the formation of
two three-dimensional operator subspaces $X^-$, $Y^-$, $Z^-$ and $X^+$, $Y^+$, $Z^+$ (the two 
subspaces are respectively zero- and double-quantum operators in a frame tilted by $\pi/2$ around the 
$I_y$ and $S_y$ axes).
 
Using this notation, the widely used Double Cross Polarization (DCP) \cite{dcp} experiment may be described by 
choosing $p-q = \pm 1$ and $|p+q| > 2$  to obtain 
\begin{eqnarray}
\label{effHam}
\bar{H}_{IS}^{(\pm 1)} &=& \kappa [ \cos(\gamma) Z^{-} \pm \sin(\gamma)  Y^{-} ] \quad ,
\end{eqnarray}where $\gamma$ as before is the 
Euler angle discriminating the crystallites by rotations around the rotor axis. The scaling factor
$\kappa = \frac{1}{2 \sqrt{2}}b_{IS} \sin(2 \beta)$ depends on the dipole-dipole coupling constant
$b_{IS}$  and the angle $\beta$ between the internuclear axis and the rotor axis. 
This effective Hamiltonian mediates the coherence transfer $I_x \rightarrow S_x$ with an efficiency
independent on the $\gamma$ Euler angle. The details of the transfer process become transparent by decomposing 
$I_x = X^{+} + X^{-}$ and observing that the effective Hamiltonian $\bar{H}_{IS}^{(\pm 1)}$, commutes with 
the operator $X^{+}$, while the operator $X^{-}$ undergoes the transformation 
\begin{equation}
\label{Eq:transf} 
    X^{-} \rightarrow  \cos(\kappa t) X^{-} + \sin(\kappa t) [\cos(\gamma) Y^{-} \mp \sin(\gamma) Z^{-}] \quad ,
\end{equation}
where the maximum transfer onto $S_x = X^{+} - X^{-}$ is obtained when $\cos(\kappa t)= -1$ \cite{gate,idcp}
(In the subsequent text $\pm 1$ in $\bar{H}_{IS}^{(\pm 1)}$ is omitted for simplicity).

If there is no dispersion in $\kappa$, evolution under the effective Hamiltonian in Eq. (\ref{effHam}) 
in $\frac{\pi}{\kappa}$ units of time leads to inversion of $X^{-}$ and makes a complete 
$I_x \rightarrow S_x$ transfer. For a powder sample, however, the dispersion in $\kappa$ has the consequence 
that for no single evolution time complete polarization transfer for all orientations can be achieved. 
The evolution time is therefore chosen for a nominal value $\kappa_0$ such that it gives the optimal polarization 
transfer of 73\% when averaged over the whole powder \cite{idcp}. For a powder sample this leads to trajectories of 
the type shown in Fig. \ref{fig:Fig1}a. The trajectories correspond to different values of $\gamma$ and a specific 
$\beta$ value that executes 86\% of a full $\pi$ rotation. The corresponding DCP pulse sequence 
is shown in Fig. \ref{fig:Fig2}a. It is clear from Fig. \ref{fig:Fig1}a that while the 
paths of the various trajectories depend on the $\gamma$ crystallite angle, the net transfer is 
independent of $\gamma$. It is also clear that,
except for the ideal $\kappa t = \pi$ rotation, there will be a loss in transfer efficiency.  

It is possible to compensate for dispersion in the value of $\beta$ 
(hence $\kappa$) by composing the evolution of the spin system under $\bar{H}_{IS}$ with supplementary 
rf rotations to 
achieve a compensating effect very similar to composite pulses in 
liquid-state NMR  \cite{levittprog}. The basic idea of 
this compensation is illustrated in 
Fig. \ref{fig:Fig1}b by a trajectory starting with an initial operator $X^{-}$ 
(the angle $\gamma$ is chosen to be 0). The parts of the trajectory labeled $I$ , 
$III$, and $V$ denote the evolution under the 
effective Hamiltonian $\bar{H}_{IS}$ for duration corresponding to nominal rotation angles
$\frac{\pi}{2}$, $\pi$, and 
$\frac{\pi}{2}$, respectively. The sections $II$ and $IV$ of the trajectory represent a 
$-\frac{\pi}{2}$ and $\frac{\pi}{2}$ angle rotation under the rf Hamiltonian $X^-$, respectively. 
A $\frac{\pi}{2}$ rotation around $X^-$ can be achieved by $x$-phase $\frac{\pi}{2}$ pulse
on the $I$- or the $S$-spin rf channel. Alternatively one may apply $\frac{\pi}{4}$ pulses on both channels. The 
compensation in dispersion of $\beta$ is immediate from the trajectory. The internuclear vectors with a larger value of 
$\kappa$ execute a 
bigger rotation during the phase $I$ of the trajectory. The compensating $\pi$ rotation (phase III) 
swaps (approximately) the position of internuclear vectors with 
values $\kappa_0 (1 + \epsilon)$ and $\kappa_0 (1 - \epsilon)$, where $\epsilon$ is the fractional 
dispersion from the nominal value $\kappa_0$. In the final phase (phase V) the larger values 
$\kappa_0 (1 + \epsilon)$ catch up with $\kappa_0 (1 - \epsilon)$ at the end point. The corresponding 
unitary evolution is described by the propagator
\begin{equation}
\label{Eq:propagator_cdcp} 
U = \exp(-i \bar{H}_{IS}t_{\frac{\pi}{2}})\exp(-i \frac{\pi}{2}
X^-) \exp(-i \bar{H}_{IS}t_{\pi})\exp(i \frac{\pi}{2}X^-) \exp(-i \bar{H}_{IS}t_{\frac{\pi}{2}})  \quad ,
\end{equation}
where $t_\frac{\pi}{2}$ and $t_\pi$ denote nominal evolution periods for $\pi/2$ and $\pi$ rotations under influence of 
$\bar{H}_{IS}$. The propagator can be experimentally realized using the pulse sequence in Fig. \ref{fig:Fig2}b. 
 The sequence comprises three evolution periods of the DCP Hamiltonian $\bar{H}_{IS}$ 
inserted into a scaffold (comb) of phase correcting pulses and will
henceforth be referred to as COMB$_3$DCP. The term COMB, proposed for future reference,
stands for compensated for beta, which is a concept which adds efficiency
as a complement to gamma encoding.

The compensated recoupling is in complete analogy with the classic $(\pi/2)_{y}(\pi)_{\bar x}(\pi/2)_{y}$ 
inversion pulse 
that is commonly used to compensate for rf inhomogeneity in liquid-state NMR  \cite{levittfreeman,levittprog}. 
In Fig. \ref{fig:Fig1}b, the $(\pi/2)_{y}$ parts of the pulse sequence are the phases $I$ and $V$ 
of the trajectory and corresponds to the evolution of $\bar{H}_{IS}$. The $(\pi)_{\bar x}$ rotation is achieved by the 
segments $II-III-IV$ and corresponds to the unitary transformation $\exp(-i \frac{\pi}{2}X^-) 
\exp(-i\bar{H}_{IS}t_{\pi})\exp(i \frac{\pi}{2}X^-)$. The $\frac{\pi}{2}$ 
rotations around $X^-$ take significantly less time compared to the evolution of $\bar{H}_{IS}$.
The methodology presented here works for all values of $\gamma$, merely leading to a rotation of the trajectory 
around the $X^-$ axis, and therefore preserves the attractive $\gamma$-encoding property \cite{horror,idcp} of the 
standard DCP experiment and in 
addition compensates for the dispersion in the $\beta$ crystallite angle. The picture in Fig. \ref{fig:Fig1}b 
presents a simple one spin 
analogy between well studied composite rf pulses in liquid-state NMR with the conceptually 
more complex coherence transfers between coupled spins under MAS conditions.  We note that similar analogy has been exploited before 
 in liquid state NMR for e.g., in ``RJCP'' compensation schemes \cite{chingas1, chingas2} (see \cite{glaser} for a review).
 This analogy immediately helps us to construct more elaborate anisotropy compensation schemes like dipolar-recoupling analogs to the 
composite pulse  $(3\pi/2)_{\bar x}(2\pi)_{x}(\pi/2)_{y}(3\pi/2)_{\bar y}(2\pi)_{y}(\pi/2)_{x}$ 
\cite{shaka}. The various phases are obtained by simply observing that the phase of rotation under 
$\bar{H}_{IS}$ can be advanced by $\theta$ by inserting the evolution under 
$\bar{H}_{IS}$ between a $-\theta$ and $\theta$ rotation around $X^-$. 
In shorthand notation the sequence may be written:
$3t_{\pi/2} - (\pi/2)^I_{\bar{x}}(\pi/2)^S_{x} - 4t_{\pi/2} - (\pi/4)^I_{\bar{x}}(\pi/4)^S_{x} 
- t_{\pi/2} - (\pi/2)^I_{\bar{x}}(\pi/2)^S_{x}
- 3 t_{\pi/2} - (\pi/2)^I_{\bar{x}}(\pi/2)^S_{x} - 4t_{\pi/2} - (\pi/4)^I_{x}(\pi/4)^S_{\bar{x}} - t_{\pi/2}$, where 
$t_{\pi/2}$ corresponds to a period with DCP-matched rf irradiation corresponding to a $\pi/2$ rotation under action of 
$\bar{H}_{IS}$. This sequence will henceforth be referred to as COMB$_6$DCP. 

The  role of rf inhomogeneity also becomes transparent in this single spin analogy to the two-spin 
coherence transfer process in Fig. \ref{fig:Fig1}b. If $\Delta \omega_{rf}^I$ and $\Delta \omega_{rf}^S$ represents 
the dispersion in the $I$- and $S$-spin  rf field  strengths from their nominal values then the effective Hamiltonian 
in Eq. (\ref{effHam}) is modified to 
\begin{equation}
\label{Eq:inhomham} 
\bar{H} = \bar{H}_{IS} + \Delta \omega_{rf}^{-} X^{-} + \Delta \omega_{rf}^{+} X^{+} \quad , 
\end{equation}
where $\Delta \omega_{rf}^{\pm}= \frac{1}{2}(\Delta \omega_{rf}^I \pm \Delta \omega_{rf}^S)$. The operator 
$X^{+}$ commutes with $\bar{H}_{IS}$ and consequently has no effect on evolution of the initial coherence $I_x$. 
In the single spin analogy, the term  $\Delta \omega_{rf}^{-} X^{-}$ produces an uncontrolled rotation around 
$X^-$ axis and acts like a Larmor 
frequency (or resonance offset) dispersion in 
Fig. \ref{fig:Fig1}b. Therefore, the anisotropies in $\beta$ and rf-field strengths translate to problems of 
rf-inhomogeneity and frequency dispersion in the single 
spin picture in Fig. \ref{fig:Fig1}. Compensating for both these dispersions in recoupling experiments is reduced to finding an inversion pulse that compensates both for rf inhomogeneity and resonance offset. The  
 $(\pi/2)_{y}(\pi)_{\bar x}(\pi/2)_{y}$ and 
 $(3\pi/2)_{\bar x}(2\pi)_{x}(\pi/2)_{y}(3\pi/2)_{\bar y}(2\pi)_{y}(\pi/2)_{x}$  
pulse schemes do provide compensation for 
both these dispersions. It is possible now to construct more elaborate compensation schemes that achieve 
a higher level of compensation.

The increased level of compensation comes at the cost of a longer pulse sequence. This naturally 
leads to the problem of finding the shortest possible pulse sequence that achieves a desired level of compensation 
for a prescribed distribution of the angle $\beta$ and inhomogeneity in rf-field strength. 
This is a problem in optimal control \cite{pontryagin} and can be addressed 
rigorously in this framework as done recently in a numerical approach \cite{ocdcp}. The analogy of dipolar recoupling to 
control of single spin reduces the problem of designing short and robust recoupling experiments in the presence of dispersion 
in $\beta$ and rf-field strength to design of short inversion pulses for a single spin that are 
robust to Larmor dispersion and rf-inhomogeneity. This problem has recently been studied in detail in 
the framework of optimal control \cite{kobzar}. It is expected 
that many of these ideas and techniques can be directly translated to design 
recoupling experiments that achieve compensation comparable to adiabatic sequences in a 
much shorter time \cite{hetadiabat,homoadiabat}. The single spin analogy also offers an explanation 
to the superior performance of numerical optimal control procedures demonstrated in solid state NMR 
recoupling experiments  \cite{ocdcp,ochorror}. 

The ideas presented above are not restricted to dipolar recoupling driven heteronuclear coherence transfer but may also 
readily be adapted to obtain anisotropy compensated homonuclear dipolar recoupling. This becomes evident  if we consider 
two dipolar coupled homonuclear spins $I$ and $S$  for which the MAS modulated dipolar-coupling 
Hamiltonian is of the form 
\begin{equation}
\label{Eq:homham}
H_{IS}(t)  = \omega_{IS}(t) (\textbf{I}.\textbf{S} - 3I_zS_z)
\end{equation}
In the interaction frame of a non-selective constant-phase rf irradiation, the $\omega_{IS}(t) \textbf{I}.\textbf{S}$ component averages to zero 
over a rotor period as the operator term $\textbf{I}.\textbf{S}$ commutes with the rf Hamiltonian (e.g., $I_x + S_x$). This leaves us with
the modulation of the $3 I_zS_z$ component, and a formalism very similar to that described 
above for the heteronuclear case
applies. The only difference is that one has non-selective rf irradiation in this case and as a result one manipulates
operators of the the type $X^+$ instead  of the differential operator $X^-$. A good example is the HORROR experiment \cite{horror} in which the two spins are irradiated at their mean resonance frequency by a rf field with the amplitude $\omega_{rf}$ adjusted to half the 
rotor frequency, i.e., $\omega_{rf} = \frac{1}{2}\omega_r$. Using Eq. (\ref{eq:effmain}) with $p=q=\frac{1}{2}$ the 
dipolar coupling Hamiltonian in the interaction frame of the rf irradiation averages over a rotor period to 
\begin{equation}
\label{Eq:homhameff} 
\bar{H}_{IS} = \kappa [\cos(\gamma) Z^{+} + \sin(\gamma)  Y^{+}],
\end{equation}
with $\kappa = \frac{3}{4 \sqrt{2}}b_{IS} \sin(2 \beta)$. In this case the effective Hamiltonian commutes with 
the operator $X^{-}$ and inverts $X^{+}$ after a time evolution 
corresponding to a $\pi$ pulse. This situation may be represented by the diagram in Fig. \ref{fig:Fig1}a except 
interchanging the
"zero-quantum" coordinate system $\{X^-,Y^-,Z^-\}$ with the corresponding "double-quantum" coordinate system
$\{X^+,Y^+,Z^+\}$. Within this framework we can compensate for dispersions in the dipolar coupling scaling factor 
$\kappa$ using the same composite sequences as used for the DCP experiment, by replacing the rotations in segments II and 
IV of the trajectories (Fig. \ref{fig:Fig1}b) by $\frac{\pi}{2}$ rotations produced by rf Hamiltonian $X^+ = \frac{I_x + S_x}{2}$. 
As described above, more elaborate compensation schemes that simultaneously compensate for the dispersion in 
$\beta$ and inhomogeneity in rf field can also be found in this case.

Finally, it is relevant to mention that the COMB dipolar recoupling is not restricted to its analogy with
liquid-state NMR composite $\pi$ inversion pulses. Compensated $\pi$/2 excitation pulses may serve as inspiration to design dipolar
recoupling experiments for which it is relevant to excite coherences at the equator of the three-dimensional operator
representations in Fig. \ref{fig:Fig1}. This may be the case, for example, in double- or zero-quantum experiments where, e.g.
double-quantum filtration or coherence evolution may provide information about spin topologies or through coupling to other
spins information about dihedral angles. One example could be the double-quantum (2Q) HORROR experiment \cite{horror} where
$\gamma$-encoded excitation and reconversion of double-quantum coherences is an intrinsic element. In this case it is possible to
maintain $\gamma$-encoding and compensate for dispersions in the dipolar coupling using composite excitation pulses such
as the classic $(\pi/2)_{x}(\pi)_{y + \pi/6}$ composite excitation pulse.

\section{Results and Discussion}

A first impression of the performance of the anisotropy compensated heteronuclear dipolar recoupling DCP experiments 
relative to the conventional DCP experiment is illustrated in Fig. \ref{fig:Fig3}a with numerical calculations of the efficiency
of a typical $^{15}$N $\rightarrow$ $^{13}$C coherence transfer as function of the excitation period. The calculations 
address specifically the 
$^{13}$C$_\alpha$-$^{15}$N spin pair of glycine in a powder sample subject to 10 kHz MAS, an external magnetic 
field corresponding to a 700 MHz (Larmor frequency for $^1$H) spectrometer, and nominal rf field strengths on the 
$^{13}$C and $^{15}$N channels of 35 and 25 kHz, respectively. These graphs reveal that the compensated schemes indeed
increases the efficiency of 70.2\% for the $\gamma$-encoded DCP experiment to 80.9\% and 87.1\% for the 
three- and six-pulse compensated schemes under conditions of homogeneous rf fields (we note the theoretical numbers for the
ideal case without chemical shielding anisotropies etc, and perfect digitization of the rotor period are slightly higher). 
This corresponds to gain factors of 1.15 and 1.24 for COMB$_3$DCP and
COMB$_6$DCP, respectively. Under 2\%  Lorentzian rf inhomogeneity the corresponding gain factors increase to 1.21 and 1.29, while they
increase to 1.38 and 1.83 for 5\%  Lorentzian rf inhomogeneity. We note that in these evaluations the effect of 5\% Lorentzian rf inhomogeneity resemble 9 - 10\% Gaussian inhomogeneity both of which being representative for the inhomogeneity of typical 
2.5 - 4 mm solid-state NMR rotors. 

It is clear from the graphs in Fig. \ref{fig:Fig3}a that composite recoupling experiments not only improve the 
transfer efficiency in terms of compensation for 
dispersions in the dipolar coupling but also provides significant improvements by compensation of rf inhomogeneity effects.
It is also clear  that the length of the pulse sequences increases substantially for the establishment of
the most efficient compensation and that a compromise has to be taken experimentally to avoid excessive loss due to relaxation.
This aspect becomes clear in Fig. \ref{fig:Fig3}b showing experimental $^{13}$C spectra of glycine obtained using a triple
resonance transfer scheme using CP for $^1$H $\rightarrow$ $^{15}$N transfer and DCP, COMB$_3$DCP or COMB$_6$DCP for $^{15}$N 
$\rightarrow$ $^{13}$C transfer (layout as in Fig. \ref{fig:Fig2}) using the same conditions as described for the numerical 
simulation. The experiments were acquired using a BRUKER AVANCE 700 MHz NMR spectrometer using a standard 2.5 mm 
triple-resonance MAS probe. The rf inhomogeneity of a full 2.5 mm MAS rotor resembles a 5\% Lorentzian shape. The experimental spectra show a gain factor 1.34 for COMB$_3$DCP relative to DCP, while no 
experimental gain was obtain using the somewhat longer COMB$_6$DCP pulse sequence. This loss of gain can most likely be
attributed to relaxation effects.

A clearer picture of the compensating effect of the composite recoupling experiments is given in Fig. \ref{fig:Fig4} by 3D
plots of the transfer efficiency as function of the rf inhomogeneity parameter $\omega_{rf}/\omega_{rf}^{nom}$ 
and the dipolar coupling deviation parameter $\omega_D/\omega_D^{nom}$ for the DCP,  
COMB$_3$DCP, and COMB$_6$DCP recoupling sequences. These plots, calculated for a powder of glycine using the same
conditions as in Fig. \ref{fig:Fig3}, reveal two interesting points. First, it is readily seen
that the composite sequences is increasingly robust towards variations in the dipolar coupling relative to the
nominal value. This implies that a larger number of the crystallites contribute efficiently to the coherence
transfer process, translating directly into improved sensitivity. Also, it is clear that the composite sequences
are somewhat more broadbanded with respect to variations in the rf field strengths, and thereby towards rf
inhomogeneity. These features explain the conclusions drawn from Fig. \ref{fig:Fig3}. The second point to note is
that these plots very closely resemble known plots for liquid-state NMR composite inversion pulses \cite{levittprog} 
where the rf inhomogeneity axis in Fig. \ref{fig:Fig4} is replaced by deviation of the resonance offset relative 
to the nominal rf field 
strength (i.e., $\Delta \omega/\omega_{rf}^{nom}$), while the dipolar coupling deviation is replaced by rf
inhomogeneity (i.e., $\omega_{rf}/\omega_{rf}^{nom}$). This observation reinforces the very close analogy
between composite recoupling sequences and composite liquid-state NMR pulses. 

The robustness of the composite recoupling experiments towards non-correlated rf inhomogeneity on the two rf channels and
variations in the resonance offsets for the two spin species is analyzed in Fig. \ref{fig:Fig5}. It is well-known that the
original DCP experiments suffer from relatively little tolerance to both parameters as  revealed by the plots in 
Fig. \ref{fig:Fig5}a. Clearly, the COMB$_3$DCP and COMB$_6$DCP pulse schemes broaden the rf field strength and offset 
ranges over which efficient dipolar recoupling and thereby coherence transfer may be accomplished. In practical applications
the improved robustness towards rf inhomogeneity is particularly important. This applies, for example, in biological solid-state
NMR where relative large sample volumes may be required to obtain a reasonable number of spins within the sample volume 
that besides the relevant molecules should contain membranes, buffers etc to ensure functional conditions. In such cases it is
important that all spins contribute to the sensitivity of the experiment independent of their location in the MAS rotor. 
In practice, it is also desirable to have a relatively broad DCP matching condition. This not only renders the 
experiment optimization, and the transfer of optimized conditions from sensitive "setup"-samples to less sensitive real 
samples, much easier but also facilitates preservation of optimal match throughout time-consuming experiments independently 
on minor variations in the tuning of the rf circuitry. Both aspects become visible in the experimental spectra in 
Fig. \ref{fig:Fig6} showing the efficiency of $^{15}$N $\rightarrow$ $^{13}$C coherence transfer in 
$^{13}$C$_\alpha$-$^{15}$N-glycine as function of the field strength on the $^{13}$C rf channel. These spectra quite clearly
reveal that the sensitivity of COMB$_3$DCP is superior to the DCP experiment and that the rf matching profiles are 
significantly broader for both compensated experiments as predicted by the theoretical/numerical analysis.

Our results clearly validate the analogy between the simple "Bloch"-picture of rotations in three-dimensional operator
coordinate systems, as commonplace for the description of liquid-state NMR composite pulses, and the features of dipolar recoupling
. This has been demonstrated on the basis of simple cross-polarization type experiments for heteronuclear
coherence transfer with the result of improved compensation of crystallite-induced dispersions in the effective dipolar coupling
and improved robustness towards rf inhomogeneity. Both elements contribute to a significant gain in the coherence transfer 
efficiency and thereby sensitivity of the experiment. From our description it should be clear that the same approach works for both
hetero- and homonuclear recoupling experiments. In a simple picture one may consider the original recoupling experiment
executed in pieces of different length interrupted by a scaffold of "phase-correcting" hard pulses. Within this picture, it is
straightforward to see that essentially all recoupling experiments may be compensated in the same fashion, with their off-set
compensating features maintained or improved. Ideally it should be possible to compensate the experiments to approach 100\%
coherence transfer for powder samples (i.e., "complete" elimination of the orientation dependency of the coherence transfer) as known from the somewhat longer adiabatic coherence transfer experiments \cite{hetadiabat,homoadiabat}. The optimum length of the recoupling experiment obviously
depends on the dipolar scaling factor of the original decoupling experiment, the desired degree of compensation, and effects from
relaxation.

\section{Conclusion}

In conclusion, we have demonstrated that analogies between composite rf pulses in liquid-state NMR spectroscopy and 
common situations for dipolar recoupling in solid-state NMR enables the establishment of general recipes for improving coherence
transfer by recoupled dipolar interactions. This improvement is ascribed to the compensation of orientation dependent dispersion
of the size of the dipolar coupling induced by variations in the angle $\beta$ between the internuclear axis and the rotor axis.
This compensation, which may be considered a supplement to the popular $\gamma$-encoding of many modern dipolar recoupling NMR
experiments, leads to substantial improvements of hetero- and homonuclear coherence transfer processes in MAS NMR spectroscopy
of powder samples. The principles of the compensated schemes is very general, implying that we foresee that composite
recoupling will find widespread applications, for example in biological solid-state NMR spectroscopy.

\newpage

\section*{Figure captions} 

\begin{figure}[p]
 \caption{}
 \vspace*{3cm}
 \centering\epsfig{file=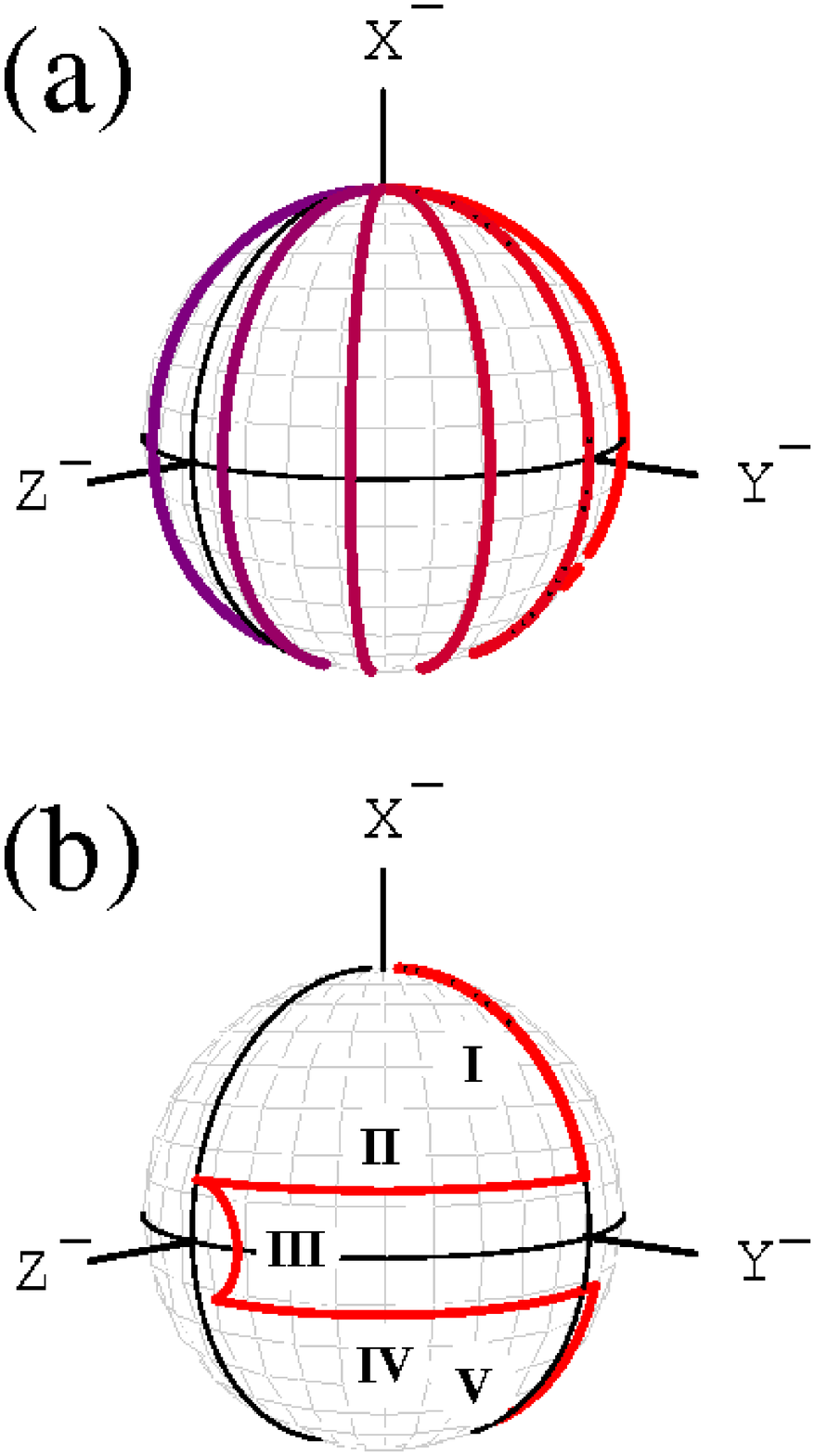, width=10cm} 
 \vspace*{3cm}
 \label{fig:Fig1}
 \end{figure}

\noindent Fig.~\ref{fig:Fig1}. Trajectories in the $X^-$, $Y^-$, $Z^-$ subspace for (a) the original DCP experiment
(different trajectories correspond to different $\gamma$ angles (from right $\gamma$=$\pi/10$ and equidistant
values decreased by $\pi/10$) as well as  (b) the $\beta$-compensated  COMB$_3$DCP experiment 
based on the combination of DCP with composite pulse procedures as described the text. The same trajectories
applies to the case of homonuclear double-quantum based coherence transfer experiments by exchanging the 
$X^-$, $Y^-$, $Z^-$ axes with $X^+$, $Y^+$, $Z^+$.

\vskip 1cm

\begin{figure}[p]
 \caption{}
 \vspace*{3cm}
 \centering\epsfig{file=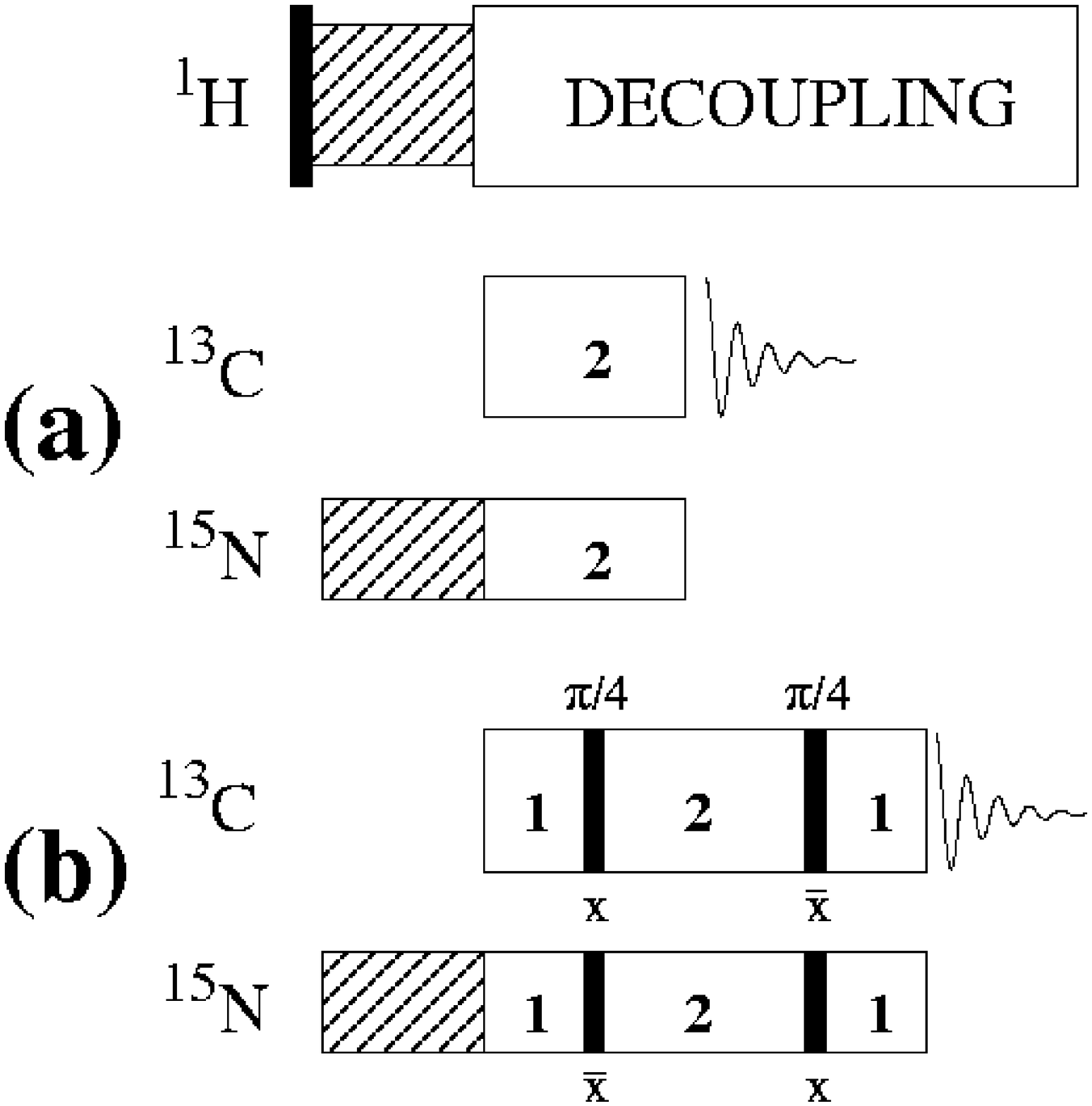, width=12cm} 
 \vspace*{3cm}
 \label{fig:Fig2}
 \end{figure}

\noindent Fig.~\ref{fig:Fig2}. Schematic representation of pulse sequences for heteronuclear coherence transfer in 
MAS solid-state NMR spectroscopy. Double Cross Polarization (DCP) pulse sequences in (a) its conventional form and (b)
with with three-element compensation (i.e., COMB$_3$DCP). The
upper trace represents $^1$H irradiation (for $^1$H $\rightarrow$ $^{15}$N coherence transfer and decoupling) applying for 
both schemes, the different height of the DCP elements (open squares) on $^{13}$C and $^{15}$N reflects mismatch in 
the rf amplitude by one
rotor frequency (e.g., $\omega_{rf}^{C} =\omega_{rf}^{N} \pm \omega_{r}$) while the number in the elements indicate the
length of the element expressed in units of $\pi/2$ rotations (i.e., in time units of $\kappa/(2 \pi)$).

\vskip 1cm


\vskip 1cm

\begin{figure}[p]
 \caption{}
 \vspace*{3cm}
 \centering\epsfig{file=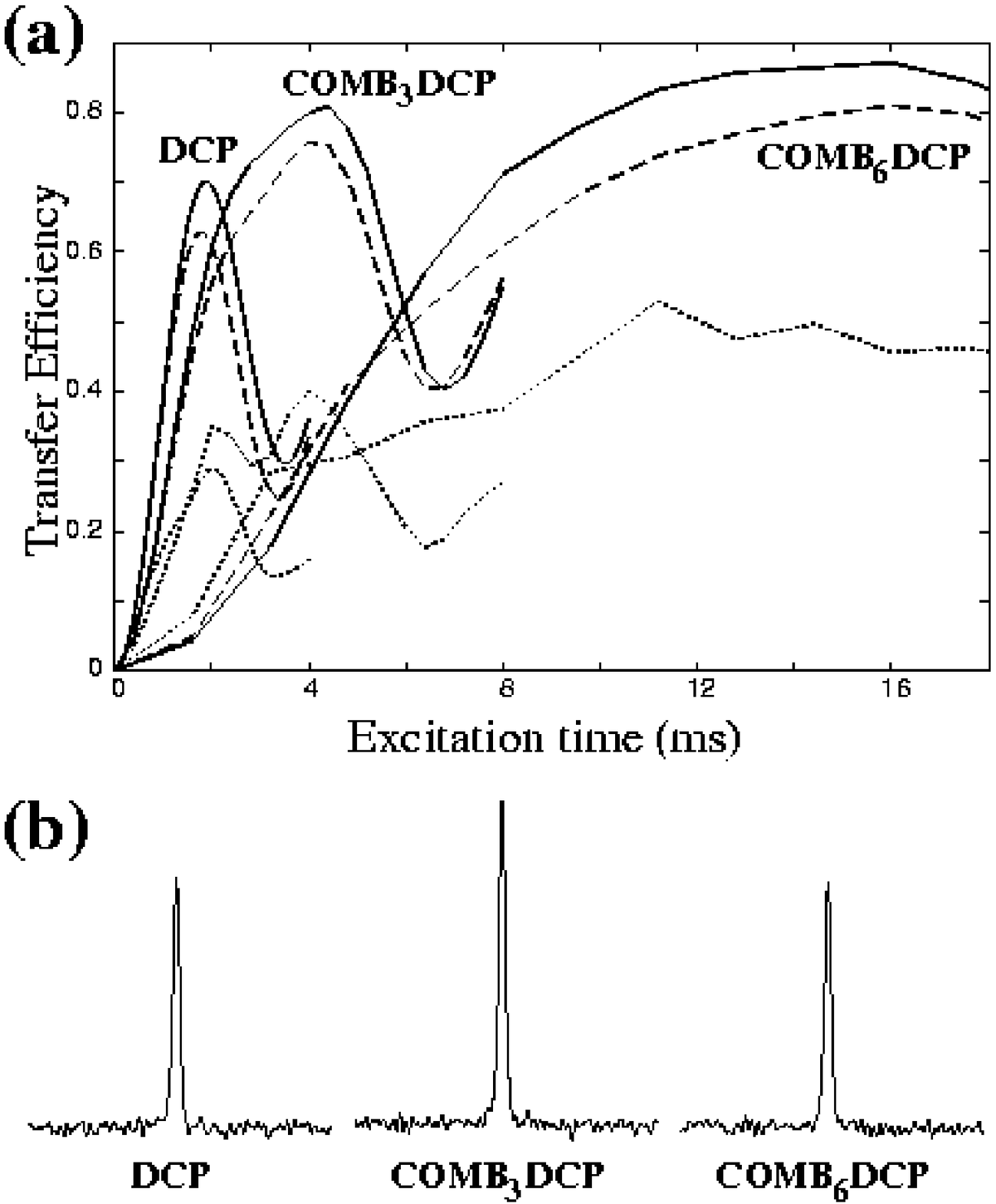, width=12cm} 
 \vspace*{3cm}
 \label{fig:Fig3}
 \end{figure}

\noindent Fig.~\ref{fig:Fig3}. (a) Efficiencies of $^{15}$N $\rightarrow$ $^{13}$C coherence transfer calculated for
DCP, COMB$_3$DCP, and COMB$_6$DCP using the parameters of glycine with experimental conditions corresponding to MAS experiments
with 10 kHz spinning and using a 700 MHz (Larmor frequency for $^1$H) magnet. The various curves reflect homogeneous 
rf field (solid line), 2\% Lorentzian rf inhomogeneity (half width at full height of the rf field distribution around 
the nominal value; dashed lined), and 5\% Lorentzian rf inhomogeneity (dotted line) on both channels using nominal 
rf field strengths of $\omega_{rf}/2\pi^C$ = 35 kHz and $\omega_{rf}/2\pi^N$ = 25 kHz. (b) Experimental spectra for DCP,
COMB$_3$DCP, and COMB$_6$DCP $^{15}$N $\rightarrow$ $^{13}$C transfers for a powder of $^{13}$C$_\alpha$,$^{15}$N-labeled 
glycine using the same conditions as used for the calculations. The experimental rf inhomogeneity 
(full 2.5 mm rotor) were estimated to be approximately 5\% Lorentzian. The calculations used $\delta_{aniso}^C$ = 19.43 ppm, $\eta^C$ = 0.98, $\{\alpha_{PR}^C,\beta_{PR}^C,\gamma_{PR}^C\}$ = $\{64.9^\circ, 37.3^\circ, -28.8^\circ\}$, 
$\delta_{aniso}^N$ = 10.1 ppm, $\eta^N$ = 0.17, $\{\alpha_{PR}^N,\beta_{PR}^N,\gamma_{PR}^N\}$ = $\{-83.8^\circ, -79.0^\circ, 
0.0^\circ\}$, $b_{CN}/2\pi$ = -890 Hz, $\{\alpha_{PR}^{CN},\beta_{PR}^{CN},\gamma_{PR}^{CN}\}$ = 
$\{0^\circ, 0^\circ, 0^\circ \}$, and $J_{CN}$ = -11 Hz.

\vskip 1cm

\begin{figure}[p]
 \caption{}
 \vspace*{3cm}
 \centering\epsfig{file=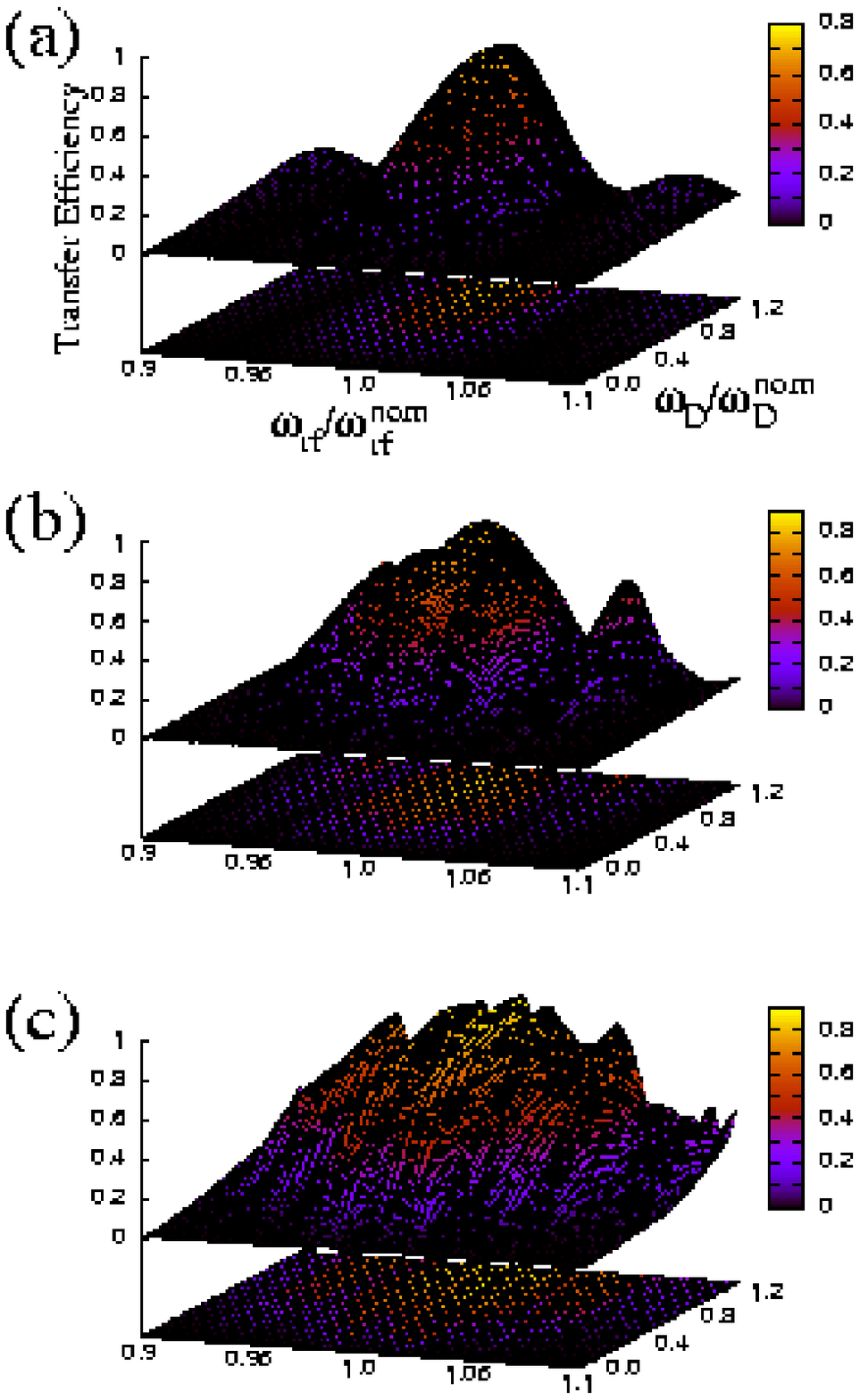, width=10cm} 
 \vspace*{3cm}
 \label{fig:Fig4}
 \end{figure}

\noindent Fig.~\ref{fig:Fig4}. 
Numerical calculations of the $^{15}$N $\rightarrow$ $^{13}$C coherence transfer for a 
powder of $^{13}$C$_\alpha$,$^{15}$N-labeled glycine (10 kHz spinning, $^1$H Larmor frequency of 700 MHz) as
function of rf inhomogeneity (expressed as the ratio between the actual rf field strength and the nominal
rf field strength, i.e., $\omega_{rf}/\omega_{rf}^{nom}$ and the deviation in the dipolar coupling 
expressed as the actual dipolar coupling relative to its nominal values, i.e., 
$\omega_D/\omega_D^{nom}$). The nominal rf field strengths were 25 kHz for $^{15}$N and 35 kHz for 
$^{13}$C (both are varied simultaneously), while the nominal dipolar coupling constant was -890 Hz (parameters as in 
Fig. \ref{fig:Fig3}).

\vskip 1cm

\begin{figure}[p]
 \caption{}
 \vspace*{3cm}
 \centering\epsfig{file=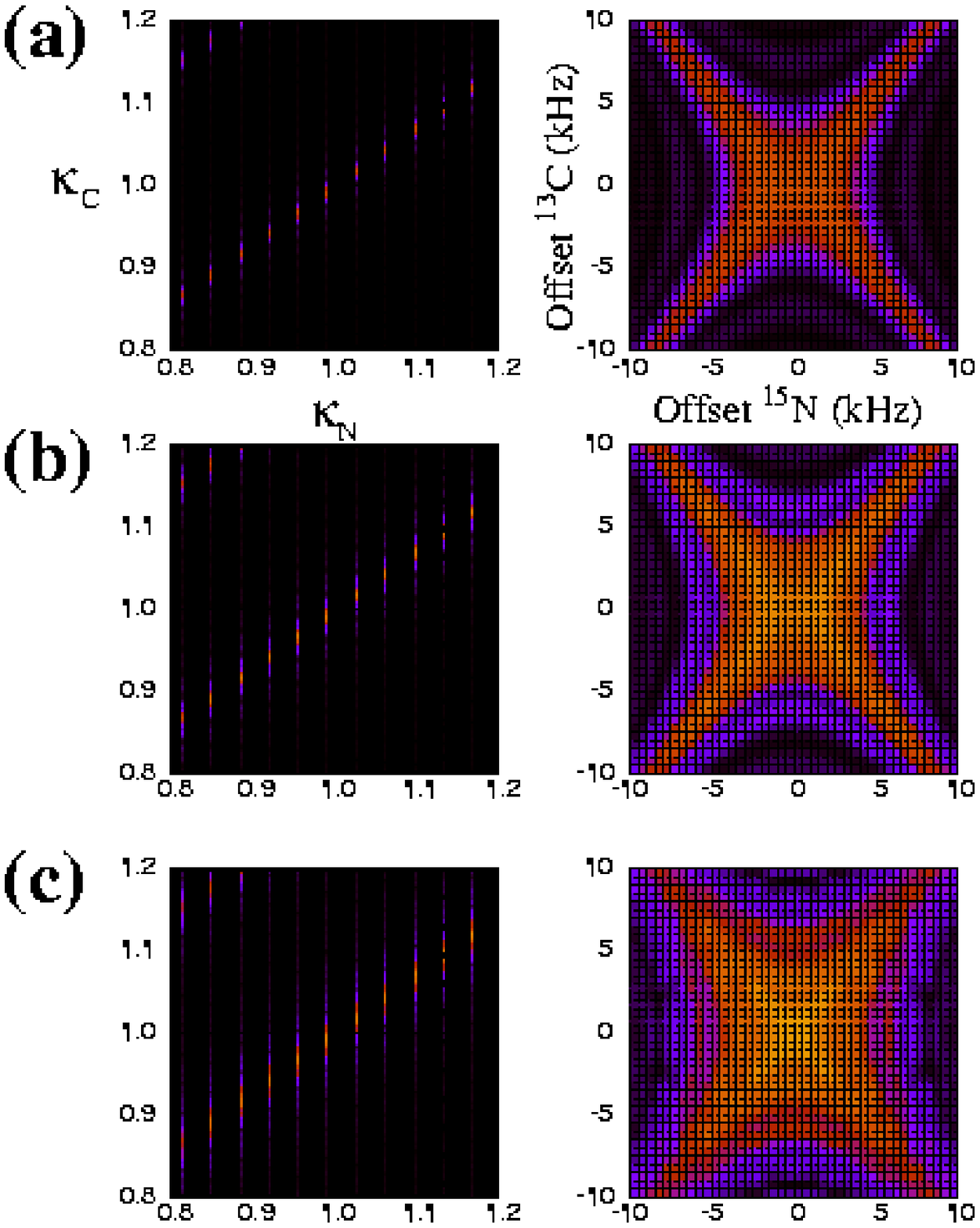, width=12cm} 
 \vspace*{3cm}
 \label{fig:Fig5}
 \end{figure}

\noindent Fig.~\ref{fig:Fig5}. Rf inhomogeneity (left column) and resonance offset (right column) profiles calculated
for (a) DCP, (b) COMB$_3$DCP, and (c) COMB$_6$DCP for a powder of glycine using the same conditions as in Fig. \ref{fig:Fig3}. 
The color coding used for the contours is identical to this used in Fig. \ref{fig:Fig4}.

\vskip 1cm

\begin{figure}[p]
 \caption{}
 \vspace*{3cm}
 \centering\epsfig{file=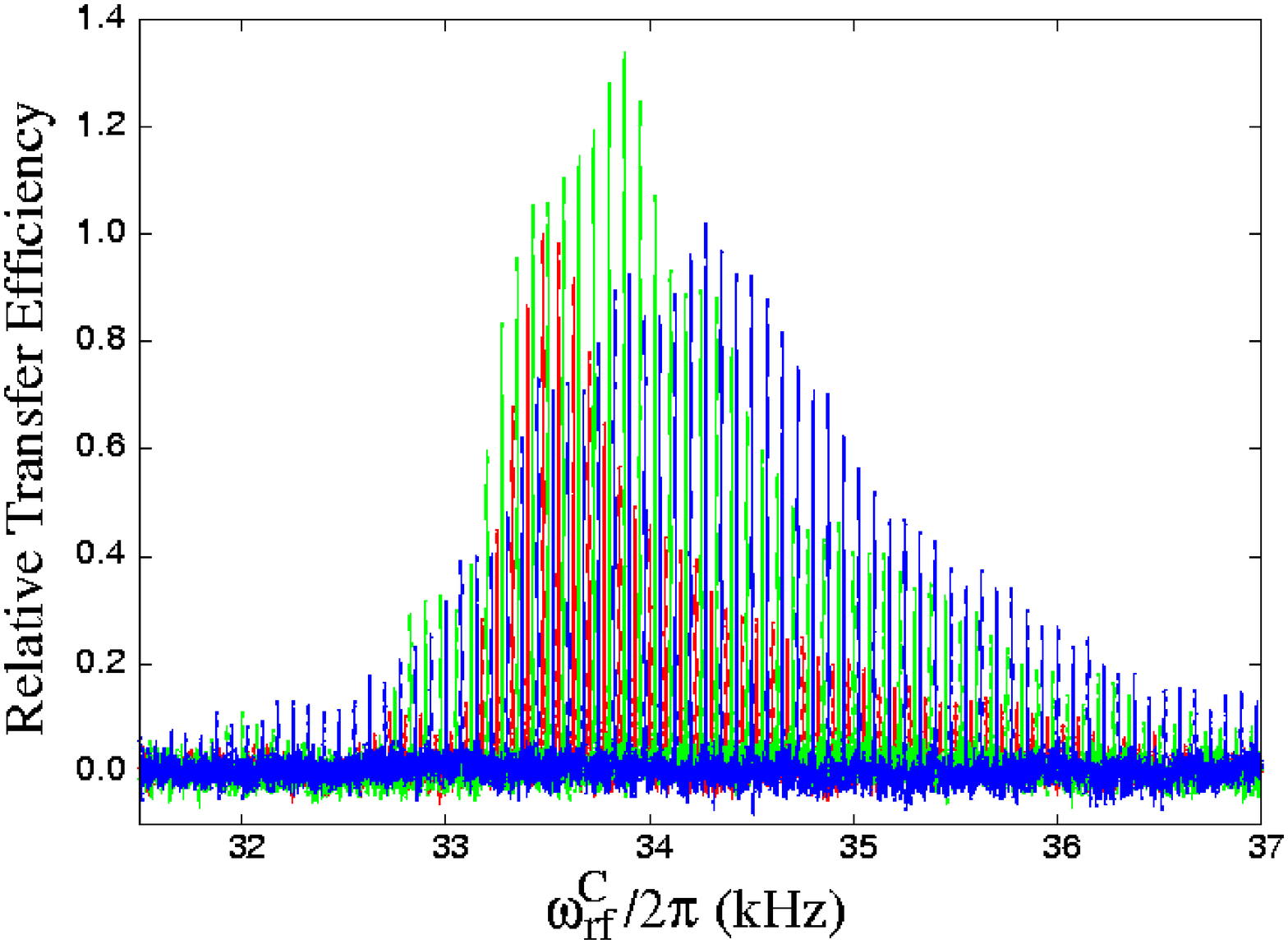, width=12cm} 
 \vspace*{3cm}
 \label{fig:Fig6}
 \end{figure}

\noindent Fig.~\ref{fig:Fig6}. 
Experimental $^{15}$N $\rightarrow$ $^{13}$C coherence transfer efficiencies for $^{13}$C$_\alpha$,$^{15}$N-labeled glycine 
(10 kHz spinning, $^1$H Larmor frequency of 700 MHz) recorded as function of the $^{13}$C rf field strength ($\omega_{rf}^C/2\pi$)
for DCP (red), COMB$_3$DCP (green), and COMB$_6$DCP (blue) dipolar recoupling experiments with total excitation times of 1.8, 
4.0, and 12.8 ms (ignoring the length of the short phase correcting pulses)
and using a $^{15}$N rf field strength of $\omega_{rf}^N/2\pi$ = 25 kHz

\newpage

\end{document}